\documentclass[12pt]{article}

\begin{document}

\title{\bf On the Physical Properties of the Plane Symmetric Self-Similar Solution}

\author{M. Sharif \thanks{msharif@math.pu.edu.pk} and
Sehar Aziz \thanks{sehar$\_$aziz@yahoo.com}\\
Department of Mathematics, University of the Punjab,\\
Quaid-e-Azam Campus, Lahore-54590, Pakistan}

\date{}

\maketitle

This paper discusses some of the physical properties of plane
symmetric self-similar solutions of the first kind (i.e.,
homothetic solutions). We are interested in calculating the
expansion, the acceleration, the rotation, the shear tensor, the
shear invariant, and the expansion rate (given by Raychaudhuri's
equation). We check these properties both in co-moving and
non-co-moving coordinates (only in the radial direction). Further,
the singularity structure of such solutions will be explored. This
analysis provides some interesting features of self-similar
solutions.

\newpage

\section*{\quad\quad\quad\quad\quad\quad\quad I. Introduction}

The similarity assumption reduces the complexity of partial
differential equations as it turns the governing partial
differential equations to relatively simple ordinary differential
equations. The Einstein field equations (EFEs)
\begin{eqnarray}
R_{ab}-\frac{1}{2}g_{ab}R=\kappa T_{ab},\quad (a,b=0,1,2,3)
\end{eqnarray}
are non-linear partial differential equations. Self-similar
solutions have been shown to be very useful in solving these set
of equations.

Self-similarity refers to the fact that the spatial distribution
of characteristics of motion remains similar to itself at all
times during the motion. Similarity solutions were first studied
in General Relativity (GR) by Cahill and Taub [1]. They assumed
that the solution was such that the dependent variables were
essentially functions of a single independent variable constructed
as a dimensionless combination of the independent variables. In
the simplest situation, a similarity solution is invariant under
the transformation $r\rightarrow ar$, $t\rightarrow at$ for any
constant $a$. Geometrically, they showed that the existence of a
similarity of first kind in this situation could be invariantly
formulated in terms of the existence of a homothetic vector. A
natural generalization of homothety, called kinematic
self-similarity, exists and is defined by the existence of a
kinematic self-similar (KSS) vector field. A KSS vector $\xi$
satisfies the following conditions:
\begin{eqnarray}
\pounds_{\xi}h_{ab}&=& 2\delta h_{ab},\\
\pounds_{\xi}u_{a}&=& \alpha u_{a},
\end{eqnarray}
where $h_{ab}$ is the projection tensor, and $\alpha$ and $\delta$
are constants. The similarity transformation is characterized by
the scale-independent ratio $\alpha/\delta$, which is known as the
similarity index.

By using a similarity index, Carter and Henriksen [2,3] defined
other kinds of self-similarity: namely, second, zeroth, and
infinite kinds. In the context of kinematic self-similarity,
homothety is considered as the first kind. Several authors have
explored KSS perfect-fluid solutions. The only barotropic equation
of state is compatible with self-similarity of the first kind is
$p=k\rho$.

Carr [4] has classified self-similar perfect-fluid solutions of
the first kind for the dust case ($k=0$). The case $0<k<1$ has
been studied by Carr and Coley [5]. Coley [6] has shown that the
Friedmann-Robertson Walker (FRW) solution is the only spherically
symmetric homothetic perfect-fluid solution in the parallel case.
McIntosh [7] has discussed a stiff fluid ($k=1$) being the only
compatible perfect-fluid with homothety in the orthogonal case.
Benoit and Coley [8] have studied analytic spherically symmetric
solutions of the EFE's coupled with a perfect-fluid and admitting
a KSS vector of the first, second, or zeroth kind.

Carr et al. [9] considered the KSS vector associated with the
critical behavior observed in the gravitational collapse of
spherically symmetric perfect fluid with equation of state
$p=k\rho$.  Carr et al. [10] further investigated the solution
space of self-similar spherically symmetric perfect-fluid models
and the physical aspects of these solutions. They combined the
state space description of the homothetic approach with the use of
the physically interesting quantities arising in the co-moving
approach. Maeda et al. [11] discussed the classification of the
spherically symmetric KSS perfect-fluid and dust solutions.
Recently, Sharif and Sehar investigated the classification of
cylindrically symmetric [12] and plane symmetric [13] KSS
perfect-fluid and dust solutions.

The existence of self-similar solutions of the first kind is
related to conservation laws and to the invariance of the problem
with respect to the group of similarity transformations of
quantities with independent dimensions. This can be characterized
in GR by the existence of a homothetic vector. Perveen [14]
classified plane symmetric Lorentzian manifolds according to their
homotheties and found different solutions admitting 5, 7, or 11
homotheties. Among these solutions, two correspond to 5
homotheties and five admit 7 homotheties. The only solution
admitting 11 homotheties is the Minkowski metric.

Recently, Sharif and Sehar explored the physical properties of
spherically symmetric self-similar solution of the first kind [15]
and cylindrically symmetric self-similar solution of the first
kind [16]. This is the third paper in the series. Here, we are
extending the same analysis for the plane symmetric self-similar
solution of the first kind. The paper can be outlined as follows:
In Section II, we shall write down the self-similar solutions of
the plane symmetric spacetime. Section III is devoted to a
discussion of the physical properties of these solutions both in
co-moving and non-co-moving coordinates. In section IV, we shall
explore the singularity structure of these solutions. Finally, we
shall summarize and discuss all the results in Section V.

\section*{II. Plane Symmetric Self-Similar Solutions of the First Kind}

The general plane symmetric spacetime is given by the line element
[17]
\begin{equation}
ds^2=e^{2\nu(t,x)}dt^2-e^{2\lambda(t,x)}dx^2
-e^{2\mu(t,x)}(dy^2+dz^2),
\end{equation}
where $\nu$, $\lambda$ and $\mu$ are arbitrary functions of $t$
and $x$. Perveen [14] classified plane symmetric Lorentzian
manifolds by homotheties and found self-similar solutions of the
first kind. There are two classes of such solutions, one admitting
5 homotheties and the other admitting 7 homotheties. This paper is
devoted to discussing the physical properties of these solutions.

The first metric is given by
\begin{equation}
ds^2=e^{2\nu(x)}dt^2- dx^2-e^{2\mu(x)}(dy^2+dz^2).
\end{equation}
This metric has the following two solutions having 5 and 7
homotheties, respectively:
\begin{equation}
ds^2=(\frac{x}{x_0})^{2A} dt^2- dx^2-(\frac{x}{x_0})^{2B}
(dy^2+dz^2),
\end{equation}
where $A\neq B,~B\neq0,~A\neq1$, and
\begin{equation}
ds^2=(\frac{x}{x_0})^{2A} (dt^2-dy^2-dz^2)- dx^2,
\end{equation}
where $A\neq0$. The second metric has the form
\begin{equation}
ds^2=dt^2-e^{2\lambda(t)} dx^2-e^{2\mu(t)}(dy^2+dz^2).
\end{equation}
This metric has the following two solutions with 5 and 7
homotheties, respectively:
\begin{equation}
ds^2=dt^2-(\frac{t}{t_0})^{2A} dx^2-(\frac{t}{t_0})^{2B}
(dy^2+dz^2),
\end{equation}
where $A\neq B,~B\neq0,~A\neq1$, and
\begin{equation}
ds^2=dt^2-(\frac{t}{t_0})^{2A} (dx^2+dy^2+dz^2),
\end{equation}
where $A\neq0$.

The metric
\begin{equation}
ds^2=e^{2f(x)}dt^2-dx^2-e^{2\frac{t}{a}+2f(x)}(dy^2+dz^2),
\end{equation}
where $a\neq0$, has only one solution admitting 7 homotheties, and
that solution is given by
\begin{equation}
ds^2=(\frac{x}{x_0})^2 dt^2- dx^2-(\frac{x}{x_0})^2
e^{2\frac{t}{a}}(dy^2+dz^2).
\end{equation}
The metric
\begin{equation}
ds^2=dt^2-e^{2f(t)}dx^2-e^{2\frac{x}{a}+2f(t)}(dy^2+dz^2),
\end{equation}
where $a\neq0$, also has one solution with 7 homotheties, and that
solution is given by
\begin{equation}
ds^2= dt^2-(\frac{t}{t_0})^2 dx^2-(\frac{t}{t_0})^2
e^{2\frac{x}{a}}(dy^2+dz^2).
\end{equation}
Finally, the metric
\begin{equation}
ds^2=dt^2-dx^2-e^{2(at+bx)}(dy^2+dz^2),
\end{equation}
yields the following solution with 7 homotheties:
\begin{equation}
ds^2= dt^2- dx^2-e^{2a(t+x)}(dy^2+dz^2).
\end{equation}

Thus, we have total of seven self-similar solutions that are given
by Eqs.(6), (7), (9), (10), (12), (14), and (16). These can be
divided into two classes admitting 5 and 7 homotheties,
respectively.

\section*{\quad\quad III. Kinematics of the Velocity Field}

This section is devoted to discussing  the kinematical properties
of the self-similar solutions of the first kind both in co-moving
and non-co-moving coordinates. These properties [17] can be listed
as follows: The volume behavior of the fluid can be determined by
the expansion scalar defined by
\begin{equation}
\Theta=u^{a}_{;a}.
\end{equation}
The acceleration can be defined as
\begin{equation}
\dot{u}_a=u_{a;b} u^b,
\end{equation}
where $u_a$ is the four-vector velocity. The rotation is given by
\begin{equation}
\omega_{ab}=u_{[a;b]}+ \dot{u}_{[a}u_{b]}.
\end{equation}
The shear tensor, which provides the distortion arising in the
fluid flow leaving the volume invariant, can be found by
\begin{equation}
\sigma_{ab}=u_{(a;b)}+\dot{u}_{(a}u_{b)} -\frac{1}{3}\Theta
h_{ab}.
\end{equation}
The shear scalar, which gives the measure of anisotropy, is
defined by
\begin{equation}
\sigma=\sigma_{ab}\sigma^{ab}.
\end{equation}
The expansion rate with respect to proper time is given by
Raychaudhuri's equation [18]
\begin{equation}
\frac{d\Theta}{d\tau}=-\frac{1}{3}\Theta^2-\sigma_{ab}\sigma^{ab}
+\omega_{ab}u^{a}u^{b}-R_{ab}u^{a}u^{b}.
\end{equation}
We now discuss these properties for the self-similar solutions
given in the previous section.

\subsection*{\quad 1. Kinematic Properties in Co-moving Coordinates}

First, we evaluate the kinematical properties of the self-similar
solutions in co-moving coordinates. For this purpose, we have
divided the solutions into two classes. Class 1 has solutions that
admit 5 homotheties while class 2 has solutions that admit 7
homotheties.

\subsubsection*{A. Class 1}

This class has two solutions given by Eqs.(6) and (9). We start
the discussion of kinematic properties for the solution given by
Eq.(6).

The expansion scalar is zero for this spacetime in co-moving
coordinates. The only non-zero component of the acceleration turns
out to be
\begin{equation}
\dot{u}_1=-\frac{A}{x}.
\end{equation}
The non-vanishing rotation component becomes
\begin{equation}
\omega_{01}=2\frac{A}{x}(\frac{x}{x_0})^A.
\end{equation}
The shear component is
\begin{equation}
\sigma_{01}=-\omega_{01}
\end{equation}
while the shear invariant becomes
\begin{equation}
\sigma=-4\frac{A^2}{x^2}=-4{\dot{u}_1}^2.
\end{equation}
With Raychaudhuri's equation, the rate of change of expansion
takes the following form:
\begin{equation}
\frac{d\Theta}{d\tau}=4\frac{A^2}{x^2}
-\frac{A(A+2B-1)}{x^2}(\frac{x}{x_0})^{2B-2A}.
\end{equation}

For the self-similar solution given by Eq.(9), the expansion
scalar is given by
\begin{equation}
\Theta=\frac{(A+2B)}{t}.
\end{equation}
The acceleration and the rotation turn out to be zero for this
metric. The shear components are given as
\begin{eqnarray}
\sigma_{11}=\frac{(2B-5A)}{3t}(\frac{t}{t_0})^{2A},\quad
\sigma_{22}=\frac{(A-4B)}{3t}(\frac{t}{t_0})^{2B}=\sigma_{33},
\end{eqnarray}
and the shear invariant becomes
\begin{equation}
\sigma=\frac{4B^2+3A^2-4AB}{t^2}.
\end{equation}
The rate of expansion takes the form
\begin{equation}
\frac{d\Theta}{d\tau}=-\frac{1}{3t^2}(7A^2+10B^2+3A+6B-8AB).
\end{equation}

\subsubsection*{B. Class 2}

Class 2 has five solutions that are given by Eqs.(7), (10), (12),
(14) and (16). In this class, for the first solution, the
expansion scalar and rotation are zero. The only acceleration
component is given by
\begin{equation}
\dot{u}_1=-\frac{A}{x}.
\end{equation}
The only component of the shear becomes
\begin{equation}
\sigma_{01}=-2\frac{A}{x}(\frac{x}{x_0})^A,
\end{equation}
and the shear invariant takes the form
\begin{equation}
\sigma=-4\frac{A^2}{x^2}.
\end{equation}
The expansion rate takes the following form:
\begin{equation}
\frac{d\Theta}{d\tau}=\frac{A}{x^2}(A+1).
\end{equation}

For the second solution, given by Eq.(10), the expansion scalar is
given by
\begin{equation}
\Theta=\frac{3A}{t}.
\end{equation}
The acceleration and the rotation are zero. The non-zero
components of the shear are
\begin{equation}
\sigma_{11}=-\frac{A}{t}(\frac{t}{t_0})^{2A}=\sigma_{22}=\sigma_{33},
\end{equation}
and the shear invariant becomes
\begin{equation}
\sigma=\frac{3A^2}{t^2}.
\end{equation}
The rate of change of expansion takes the form
\begin{equation}
\frac{d\Theta}{d\tau}=-\frac{3A}{t^2}(A+1).
\end{equation}

The third solution in class 2 has the following expansion scalar:
\begin{equation}
\Theta=\frac{2x_0}{ax}.
\end{equation}
The only acceleration component is given by
\begin{equation}
\dot{u}_1=\frac{1}{x}
\end{equation}
while the rotation turns out to be zero. The non-vanishing
components of the shear are
\begin{eqnarray}
\sigma_{11}=\frac{2x_0}{3ax},\quad
\sigma_{22}=-\frac{4x}{3ax_0}e^{2\frac{t}{a}}=\sigma_{33},
\end{eqnarray}
and the shear invariant becomes
\begin{equation}
\sigma=\frac{4{x_0}^2}{a^2x^2}.
\end{equation}
The rate of change of expansion has the following form:
\begin{equation}
\frac{d\Theta}{d\tau}=-\frac{2}{3a^2x^2}(5{x_0}^2+3a^2).
\end{equation}

For the fourth solution, given by Eq.(14), the expansion scalar is
given by
\begin{equation}
\Theta=\frac{1}{t}.
\end{equation}
The acceleration and the rotation are zero for this metric. The
components of the shear are
\begin{equation}
\sigma_{22}=-\frac{5t}{3t^2_0}e^{2\frac{x}{a}}=\sigma_{33}
\end{equation}
while the shear invariant becomes
\begin{equation}
\sigma=\frac{50}{9t^2}.
\end{equation}
The expansion rate is given, by using Raychaudhuri's equation, as
\begin{equation}
\frac{d\Theta}{d\tau}=-\frac{53}{9t^2}.
\end{equation}

The last solution in this class, given by Eq.(16), has the
expansion scalar
\begin{equation}
\Theta=2a.
\end{equation}
The acceleration and the rotation become zero. The components of
the shear are
\begin{equation}
\sigma_{22}=-\frac{4a}{3}e^{2a(t+x)}=\sigma_{33},
\end{equation}
and the shear invariant becomes
\begin{equation}
\sigma=\frac{32a^2}{9}.
\end{equation}
The expansion rate is
\begin{equation}
\frac{d\Theta}{d\tau}=-\frac{26a^2}{9}.
\end{equation}

\subsection*{2. Kinematic Properties in Non-co-moving Coordinates}

This section is devoted to discussing the same properties of the
self-similar solutions in non-co-moving coordinates only in the
radial direction.

\subsubsection*{A. Class 1}

For the first metric in class 1, given by Eq.(6), we obtain
non-zero expansion as follows:
\begin{equation}
\Theta=-\frac{(A+2B)}{x}.
\end{equation}
The non-zero components of the acceleration turn out to be
\begin{equation}
\dot{u}_0=\frac{A}{x}(\frac{x}{x_0})^A,\quad
\dot{u}_1=-\frac{A}{x}.
\end{equation}
The non-zero rotation component is
\begin{equation}
\omega_{01}=\frac{A}{x}(\frac{x}{x_0})^A=\dot{u}_0.
\end{equation}
The components of the shear are
\begin{eqnarray}
\sigma_{00}=4\frac{A}{x}(\frac{x}{x_0})^{2A},\quad
\sigma_{01}=\frac{2}{3x}(\frac{x}{x_0})^A (B-4A),\nonumber\\
\sigma_{11}=\frac{4}{3x}(A-B),\quad
\sigma_{22}=-\frac{1}{3x}(\frac{x}{x_0})^{2B}(A+8B)=\sigma_{33},
\end{eqnarray}
and the measure of anisotropy is given by
\begin{equation}
\sigma=\frac{2}{9x^2}(49A^2+16AB+70B^2).
\end{equation}
The rate of change of expansion using Raychaudhuri's equation
becomes
\begin{equation}
\frac{d\Theta}{d\tau}=-\frac{1}{9x^2}(101A^2+134B^2+62AB-18B)-\frac{A}{x}.
\end{equation}

For the second metric in this class, we obtain non-zero expansion
as follows:
\begin{equation}
\Theta=\frac{A+2B}{t}.
\end{equation}
The non-zero components of acceleration are
\begin{equation}
\dot{u}_0=-\frac{A}{t},\quad
\dot{u}_1=\frac{A}{t}(\frac{t}{t_0})^A,
\end{equation}
and the non-zero rotation component is
\begin{equation}
\omega_{01}=\frac{A}{t}(\frac{t}{t_0})^A=\dot{u}_1.
\end{equation}
The non-vanishing components of the shear are
\begin{eqnarray}
\sigma_{00}=-\frac{2A}{t},\quad
\sigma_{01}=\frac{(10A+2B)}{3t}(\frac{t}{t_0})^A,\nonumber\\
\sigma_{11}=-\frac{(10A-4B)}{3t}(\frac{t}{t_0})^{2A},\quad
\sigma_{22}=\frac{(A-4B)}{3t}(\frac{t}{t_0})^{2B} =\sigma_{33},
\end{eqnarray}
and the shear invariant is given by
\begin{equation}
\sigma=\frac{2}{9t^2}(19A^2-68AB+22B^2).
\end{equation}
The expansion rate becomes
\begin{equation}
\frac{d\Theta}{d\tau}=-\frac{1}{9t^2}(41A^2+38B^2-106AB+18B)-\frac{A}{t}.
\end{equation}

\subsubsection*{B. Class 2}

For the first metric in this class, we obtain non-zero expansion
as follows:
\begin{equation}
\Theta=-\frac{3A}{x}.
\end{equation}
The non-zero components of the acceleration turn out to be
\begin{equation}
\dot{u}_0=\frac{A}{x}(\frac{x}{x_0})^A,\quad
\dot{u}_1=-\frac{A}{x},
\end{equation}
and the non-zero rotation component is
\begin{equation}
\omega_{01}=\frac{A}{x}(\frac{x}{x_0})^A=\dot{u}_0.
\end{equation}
The non-vanishing components of the shear are
\begin{eqnarray}
\sigma_{00}=4\frac{A}{x}(\frac{x}{x_0})^{2A},\quad
\sigma_{01}=-\frac{2A}{x}(\frac{x}{x_0})^A,\quad
\sigma_{22}=-\frac{3A}{x}(\frac{x}{x_0})^{2A}=\sigma_{33}
\end{eqnarray}
while the measure of anisotropy is given by
\begin{equation}
\sigma=\frac{30A^2}{x^2}.
\end{equation}
The rate of change of expansion becomes
\begin{equation}
\frac{d\Theta}{d\tau}=-\frac{A}{x^2}(33A+2)-\frac{A}{x}.
\end{equation}

The second solution in this class gives the non-zero expansion
\begin{equation}
\Theta=\frac{3A}{t}.
\end{equation}
The non-zero components of the acceleration are
\begin{equation}
\dot{u}_0=-\frac{A}{t},\quad
\dot{u}_1=\frac{A}{t}(\frac{t}{t_0})^A,
\end{equation}
and the non-zero rotation component is
\begin{equation}
\omega_{01}=\frac{A}{t}(\frac{t}{t_0})^A=\dot{u}_1.
\end{equation}
The components of the shear become
\begin{eqnarray}
\sigma_{00}=-\frac{2A}{t},\quad
\sigma_{01}=\frac{4A}{t}(\frac{t}{t_0})^A,\nonumber\\
\sigma_{11}=-\frac{2A}{t}(\frac{t}{t_0})^{2A},\quad
\sigma_{22}=-\frac{A}{t}(\frac{t}{t_0})^{2A} =\sigma_{33},
\end{eqnarray}
and the shear invariant is given by
\begin{equation}
\sigma=-\frac{6A^2}{t^2}.
\end{equation}
The rate of expansion becomes
\begin{equation}
\frac{d\Theta}{d\tau}=-\frac{A}{t}+\frac{A}{t^2}(3A-2).
\end{equation}

For the third solution, the non-zero expansion is given by
\begin{equation}
\Theta=-\frac{3}{x}+\frac{2x_0}{ax}.
\end{equation}
The components of the acceleration are
\begin{equation}
\dot{u}_0=\frac{1}{x_0},\quad \dot{u}_1=-\frac{1}{x},
\end{equation}
and the rotation component is
\begin{equation}
\omega_{01}=\frac{1}{x_0}=\dot{u}_0.
\end{equation}
The components of the shear take the form
\begin{eqnarray}
\sigma_{00}=\frac{4x}{{x_0}^2},\quad
\sigma_{11}=\frac{4x_0}{3ax},\quad
\sigma_{01}=-\frac{2}{3ax_0}(3a+x_0),\nonumber\\
\sigma_{22}=-\frac{x}{3a{x_0}^2}(9a+4x_0)e^{2\frac{t}{a}}=\sigma_{33}
\end{eqnarray}
while the measure of anisotropy is
\begin{equation}
\sigma=\frac{2}{9a^2x^2}(135a^2+22{x_0}^2+60ax_0).
\end{equation}
The expansion rate becomes
\begin{equation}
\frac{d\Theta}{d\tau}=-\frac{1}{9a^2x^2}(261a^2-14{x_0}^2
-156ax_0)-\frac{1}{x}+\frac{2}{a^2{x_0}^2}(a^2-{x_0}^2)e^{2\frac{t}{a}}.
\end{equation}

The fourth self-similar solution (Eq.(14)) gives the expansion as
follows:
\begin{equation}
\Theta=\frac{3a-2t_0}{at}.
\end{equation}
The acceleration components are
\begin{equation}
\dot{u}_0=-\frac{1}{t},\quad \dot{u}_1=\frac{1}{t_0}
\end{equation}
while the non-zero rotation component is
\begin{equation}
\omega_{01}=-\frac{1}{t_0}=-\dot{u}_1.
\end{equation}
We obtain the following non-vanishing components of shear:
\begin{eqnarray}
\sigma_{00}=-\frac{2}{t},\quad
\sigma_{11}=-\frac{2t}{3a{t_0}^2}(3a+2t_0),\quad
\sigma_{01}=\frac{2}{at_0}(a+t_0),\nonumber\\
\sigma_{22}=-\frac{t}{3a{t_0}^2}e^{2\frac{x}{a}}(3a+8t_0)=\sigma_{33},
\end{eqnarray}
and the shear invariant is
\begin{equation}
\sigma=\frac{1}{9a^2t^2}(45a^2+44t^2_0+24at_0).
\end{equation}
The expansion rate becomes
\begin{equation}
\frac{d\Theta}{d\tau}=\frac{1}{t}-\frac{2}{9a^2t^2}(45a^2+19t^2_0-6at_0).
\end{equation}

The expansion, the acceleration, and the rotation vanish for the
last solution in this class. The shear components are
\begin{equation}
\sigma_{22}=-4ae^{2a(t+x)}=\sigma_{33},
\end{equation}
and the shear invariant is
\begin{equation}
\sigma=32a^2.
\end{equation}
The rate of expansion becomes
\begin{equation}
\frac{d\Theta}{d\tau}=-30a^2.
\end{equation}

\section*{\quad\quad\quad\quad\quad\quad IV. Singularities}

In this section, we shall explore the singularities of the
self-similar solutions in classes 1 and 2. The Kretschmann scalar
is defined by
\begin{equation}
K=R_{abcd}R^{abcd},
\end{equation}
where $R_{abcd}$ is the Riemann tensor. For the solution given by
Eq.(6), Eq.(92) reduces to
\begin{equation}
K=\frac{2}{x^4}[A^2(A^2+1+2B^2-2A)+B^2(3B^2+2-4B)].
\end{equation}
It is clear that $K$ diverges at $x=0$. It follows that the
solution is singular at $x=0$.

For the solution given by Eq.(7), the Kretschmann scalar becomes
\begin{equation}
K=\frac{2A^2}{x^4}(6A^2+3-2A),
\end{equation}
which shows that the spacetime singularity lies at $x=0$.

The solution given by Eq.(9) has the Kretschmann scalar
\begin{equation}
K=\frac{2}{t^4}(A^2(A-1)^2+2B^2(B-1)^2+2A^2B^2+B^4).
\end{equation}
From here, it follows that the spacetime is singular at $t=0$.

For the metric given by Eq.(10), the Kretschmann scalar reduces to
\begin{equation}
K=\frac{6A^2}{t^4}(2A^2+1-2A).
\end{equation}
It is clear that $K$ diverges at $t=0$. Hence, the solution is
singular at $t=0$.

For the solution given by Eq.(12), we obtain
\begin{equation}
K=\frac{6}{a^4 x^4}(a^2-{x_0}^2)^2,
\end{equation}
which gives the spacetime singularity at $x=0$. The solution given
by Eq.(14) has the following Kretschmann scalar:
\begin{equation}
K=\frac{6}{a^2t^4}(t^2_0-a^2)^2.
\end{equation}
It follows that the spacetime singularity lies at $t=0$.

For the last solution, given by Eq.(16), the Kretschmann scalar
reduces to
\begin{equation}
K=4a^4.
\end{equation}
It turns out to be constant, which shows that this solution is
singularity free.

\section*{\quad\quad\quad\quad\quad\quad\quad V. Conclusion}

Self-similar solutions in GR are very important, and discussing
their physical features is interesting. Keeping this point in
mind, we explored some kinematic properties and the singularity
features of such solutions representing plane symmetric spacetime.
We discussed the properties both in co-moving and non-co-moving
coordinates (only in the radial direction). This provided a
comparison of these properties in the two coordinates. We explored
the acceleration, the expansion, the rotation, the shear, the rate
of change of expansion, and finally the singularity.

First, we discussed the class 1 which has two solutions. For the
first solution, given by Eq.(6), we found zero expansion in
co-moving coordinates and positive/negative expansion in
non-co-moving coordinates depending upon the values of the
constants $A$ and $B$. The acceleration and the rotation had only
one component in co-moving coordinates while two components of the
acceleration existed in non-co-moving coordinates. The rotation
component remained the same in both coordinates, except for a
factor of one-half in non-co-moving coordinates. We found the
shear invariant to be negative in co-moving coordinates and
positive/negative in non-co-moving coordinates, depending on the
values of the constants. The rate of change of expansion could be
positive/negative, depending upon the values of $A$ and $B$ in
co-moving coordinates and negative in non-co-moving coordinates.
For the second solution, given by Eq.(9), we obtained the same
expansions in both coordinates, which could be positive/negative.
These solutions had vanishing acceleration and rotation in
co-moving coordinates and one rotation and two acceleration
components in non-co-moving coordinates. The shear invariant was
positive in co-moving coordinates whereas the rate of expansion
was negative. In non-co-moving coordinates, these quantities could
be positive/negative, depending on the values of constants.

Now, we discuss the solutions of class 2. We notice that in
co-moving coordinates, the solutions given by Eqs.(7) and (12)
have non-zero acceleration component while the remaining solutions
have zero acceleration. In non-co-moving coordinates, all
solutions have non-zero acceleration components. Also, all
solutions have zero rotation in co-moving coordinates while only
one solution (Eq.(16)) has zero rotation in non-co-moving
coordinates. The solution in Eq.(7) has zero expansion in
co-moving coordinates whereas all other solutions have
positive/negative expansion. In non-co-moving coordinates, the
expansion is positive/negative for the solutions in Eqs.(7), (10),
(12), and (14) and zero for the solution in Eq.(16).

The shear invariant is negative in co-moving coordinates and
positive in non-co-moving coordinates for the solution in Eq.(7).
The expansion rate is positive in co-moving coordinates, but it is
negative in non-co-moving coordinates. For the solution in
Eq.(10), we have a positive shear scalar in co-moving coordinates
and a negative one in non-co-moving coordinates. The rate of
expansion is negative in co-moving coordinates and positive in
non-co-moving coordinates. For the solution in Eq.(12), the shear
invariant is positive in both coordinates, and the expansion rate
is negative in both coordinates. The shear scalar is positive and
the rate of expansion is negative in both coordinates for the
solution in Eq.(14). For the solution in Eq.(16), the shear
invariant is positive, and the expansion rate is negative in both
coordinates.

Finally, we discuss the singularity structure for these solutions.
The solutions given by Eqs.(6), (7), and (12), are singular at
$x=0$ while the solutions  given by Eqs.(9), (10), and (14) are
singular at $t=0$. The solution in Eq.(16) is singularity free.

We have noticed from the above discussion that the kinematic
quantities are relatively simple in co-moving coordinates as
compared to those in non-co-moving coordinates. It is worth
mentioning that the expansion of the solution in Eq.(16) turns out
to be positive in co-moving coordinates, but it is zero in
non-co-moving coordinates.

\vspace{2cm}
\begin{description}
\item  {\quad\quad\quad\quad\quad\quad\quad\quad\quad
\bf ACKNOWLEDGMENTS}
\end{description}

One of us (SA) acknowledges the enabling role of the Higher
Education Commission Islamabad, Pakistan, and appreciates its
financial support through the {\it Merit Scholarship Scheme for
Ph.D. Studies in Science and Technology (200 Scholarships)}. We
should thank Mr. Tariq Ismaeel who brought some important
literature to us.

\newpage

{\quad\quad\quad\quad\quad\quad\quad\quad\quad \bf REFERENCES}

\begin{description}

\item{[1]} M.E. Cahill and A.H. Taub, Commun. Math. Phys.
           {\bf21}, 1(1971).

\item{[2]} B. Carter and R.N. Henriksen, Annales De Physique
           {\bf14}, 47(1989).

\item{[3]} B. Carter and R.N. Henriksen, J. Math. Phys.
           {\bf32}, 2580(1991).

\item{[4]} B.J. Carr, Phys. Rev. {\bf D62}, 044022(2000).

\item{[5]} B.J. Carr and A.A. Coley, Phys. Rev. {\bf D62}, 044023(2000).

\item{[6]} A.A. Coley, Class. Quant. Grav. {\bf14}, 87(1997).

\item{[7]} C.B.G. McIntosh, Gen. Relat. Gravit. {\bf7}, 199(1975).

\item{[8]} P.M. Benoit and A.A. Coley, Class. Quant. Grav.
           {\bf15}, 2397(1998).

\item{[9]} B.J. Carr, A.A. Coley, M. Golaith, U.S. Nilsson and C. Uggla,
           Class. Quant. Grav. {\bf18}, 303(2001).

\item{[10]} B.J. Carr, A.A. Coley, M. Golaith, U.S. Nilsson and C. Uggla,
            Phys. Rev. {\bf D61}, 081502(2000).

\item{[11]} H. Maeda, T. Harada, H. Iguchi and N. Okuyama, Prog. Theor. Phys.
            {\bf108}, 819(2002); ibid {\bf 110}, 25(2003).

\item{[12]} M. Sharif and Sehar Aziz, Int. J. Mod. Phys. {\bf D14}, 1527(2005).

\item{[13]} M. Sharif and Sehar Aziz, submitted for publication.

\item{[14]} Sadia Perveen, M.Phil. Dissertation (Quaid-i-Azam University
            Islamabad, 2003).

\item{[15]} M. Sharif and Sehar Aziz, Int. J. Mod. Phys. {\bf D14}, 73(2005).

\item{[16]} M. Sharif and Sehar Aziz, Int. J. Mod. Phys. {\bf A}(2005)
           (arXiv:gr-qc/0504102).

\item{[17]} H. Stephani, D. Kramer, M. Maccallum, C. Hoenselaers and E. Herlt,
            \textit{Exact Solutions of Einstein's Field Equations}
            (Cambridge University Press, 2003).

\item{[18]} R.M. Wald, \textit{General Relativity} (University of Chicago,
            Chicago, 1984).

\end{description}

\end{document}